\documentclass[aps,pre,reprint,superscriptaddress,nofootinbib]{revtex4-2}

\usepackage{amsmath,amssymb}
\usepackage{graphicx}
\usepackage{booktabs}
\usepackage{bm}
\usepackage{microtype}
\usepackage{hyperref}
\usepackage{comment}

\graphicspath{{figures/}}

\usepackage{xcolor}

\begin{document}

\title{Slow structures, fast degrees of freedom, \\and the morphology of chess openings}

\author{Marc Barthelemy}
\email{marc.barthelemy@ipht.fr}
\affiliation{Universit\'e Paris-Saclay, CEA, CNRS, Institut de Physique Th\'eorique,
91191 Gif-sur-Yvette, France}
\affiliation{Centre d'Analyse et de Math\'ematique Sociales (CNRS/EHESS), Paris, France}
\affiliation{Complexity Science Hub, Vienna, Austria}

\author{Manlio de Domenico}
\affiliation{Department of Physics and Astronomy “Galileo Galilei,” University of Padua, Padua, Italy}
\affiliation{Istituto Nazionale di Fisica Nucleare, Sez. Padova, Padua, Italy}

\date{\today}

\begin{abstract}
Chess openings naturally separate into slow and fast degrees of
freedom: pawns evolve through irreversible moves, whereas pieces
can be rapidly reorganized within the resulting structure.
We exploit this separation to construct a quantitative morphology
of opening space. From a catalogue of 3810 named variations,
1749 reach a common depth of $N=10$ plies, corresponding to
619 distinct positions and only 435 joint pawn structures.
We characterize these positions through observables measuring
pawn commitment and spatial organization, blockade and tension,
and piece activity and centralization. Their partial independence
reveals a multidimensional opening space and in order to explore its structural organization, we construct a network
whose nodes are distinct White pawn structures, connected when
a single pawn move or capture transforms one into another.
Remarkably, 139 of the 149 observed structures belong to a
single connected component containing $98\%$ of all positions.
This network exhibits a statistically significant modular
organization, with eight communities whose representative
structures correspond to familiar pawn formations.
Rather than forming isolated categories, these formations
emerge as distinct regions of a connected structural landscape.
Our approach thus complements the traditional opening taxonomy
by revealing similarities between differently named openings
and providing a quantitative basis for comparing the positions
they produce. 

\end{abstract}

\maketitle

\section{Introduction}
\label{sec:intro}

A central problem in statistical physics and complex-systems science is
to identify effective variables that describe a system at a useful scale
without retaining all of its microscopic information
\cite{Simon1962,Kadanoff1966,Anderson1972}. Such a reduction is
particularly interesting in sequential systems, where present states
retain traces of previous decisions and where different degrees of
freedom evolve on different persistence timescales.

Chess provides a formidable setting for a quantitative study of
complex sequential systems \cite{Levy1988ComputerChessCompendium,Shannon1950}, because its rules are fixed, the state space is large but precisely defined, and extensive databases make both trajectories and
positions directly observable. One advantage is that, unlike several empirical complex systems, no hidden variables or measurement errors are required to define the microscopic configuration. In fact, a growing literature has applied
ideas from statistical physics, complex networks, and information theory
to chess \cite{Perotti2013InnovationChess,DeMarzo2023ComplexityOpenings,Blasius2009ZipfOpenings,Maslov2009PowerLawsChess,Ribeiro2013MoveByMove,Schaigorodsky2014MemoryCorrelations,Schaigorodsky2016MemoryEffects,scharnagl2004fischer,Sigman2010RapidChessRT,Atashpendar2016SequencingChess,Chowdhary2023QuantifyingPerformance,Chacoma2024EmergentComplexity,BarthelemyFragility2025,BarthelemyEntropy2025,BarthelemyChess9602026}.

Here, we focus on chess openings, traditionally organized into a historically accumulated hierarchy of names and move sequences. This taxonomy is useful for communication but does not provide a natural coordinate system for the space of positions. Broad labels such as the Sicilian Defense contain many structurally different systems, whereas other names refer to specific move sequences. Moreover, different move orders can lead to the same position. The correspondence between names, move sequences, and positions is therefore not one-to-one. Already in 1925, R\'eti \cite{Reti1998Cours} questioned the scientific foundations of opening theory, arguing for an understanding based on general principles rather than the accumulation of variations.

From a complex-systems perspective, it is interesting to (i) understand if the opening catalogue be reduced to a small set of interpretable macroscopic observables, and (ii) whether, from a chess perspective, these same
observables can tell a player what kind of game an opening tends to produce.

Interestingly, the chess rules suggest a natural organizing principle, because pawn motion is irreversible: pawns move forward but cannot retreat. Every pawn advance
therefore records a structural decision and progressively modifies the
geometry available to the other pieces. This persistence of pawn structure
has long been central to positional chess theory, from Steinitz's emphasis
on lasting pawn weaknesses and structural advantages
\cite{Steinitz1889} to modern classifications of recurring pawn formations
\cite{Soltis1995,FloresRios2015}. Knights, bishops, rooks, and queens can,
by contrast, be rerouted and redeployed much more rapidly. This suggests a
natural separation between a \emph{slow, history-bearing sector}, defined
primarily by the pawns, and a \emph{fast, rearrangeable sector}, associated
with piece organization. The separation is both dynamical and structural:
the pawns form a slowly evolving background on which the faster piece
degrees of freedom reorganize.

Here we adopt a position-based viewpoint
for opening theory and ask whether opening positions admit a simple
structural organization in terms of these slow and fast degrees of freedom. 
We first analyze these two sectors separately. For the pawns, we measure
how much irreversible advancement has occurred, where it is located, and
how the black and white pawn structures interact. We then examine the geometry of the
set of realized pawn skeletons itself. For the pieces, we measure total
activity and where that activity is directed. We finally combine the two
sectors into a morphological space of openings and use it to build
player-facing scorecards and similarity measures. This approach complements the traditional classification of chess openings by revealing structural similarities and differences. Opening
names remain useful labels, but the underlying positions form a
continuous, measurable state space in which historically distant lines
can be structurally close, and lines sharing the same family name can be
very different.

\section{From opening lines to positions}
\label{sec:data}


We use the open \texttt{lichess-org/chess-openings} 
Encyclopaedia of Chess Openings (ECO) catalogue
\cite{LichessOpenings}. It contains 3810 named opening variations, each
specified by a PGN move sequence. Note that these are definitions of opening theory, not samples of played games: no information about game frequency,
player rating, result, or engine evaluation enters this static analysis.

Rows are de-duplicated using the triple consisting of ECO code, opening name, and PGN sequence.


To compare positions at the same stage of development, every line is
truncated to a common depth of $N$ plies. We use $N=10$, so each player has made five moves.

\begin{figure*}[t]
\centering
\includegraphics[width=0.96\textwidth]{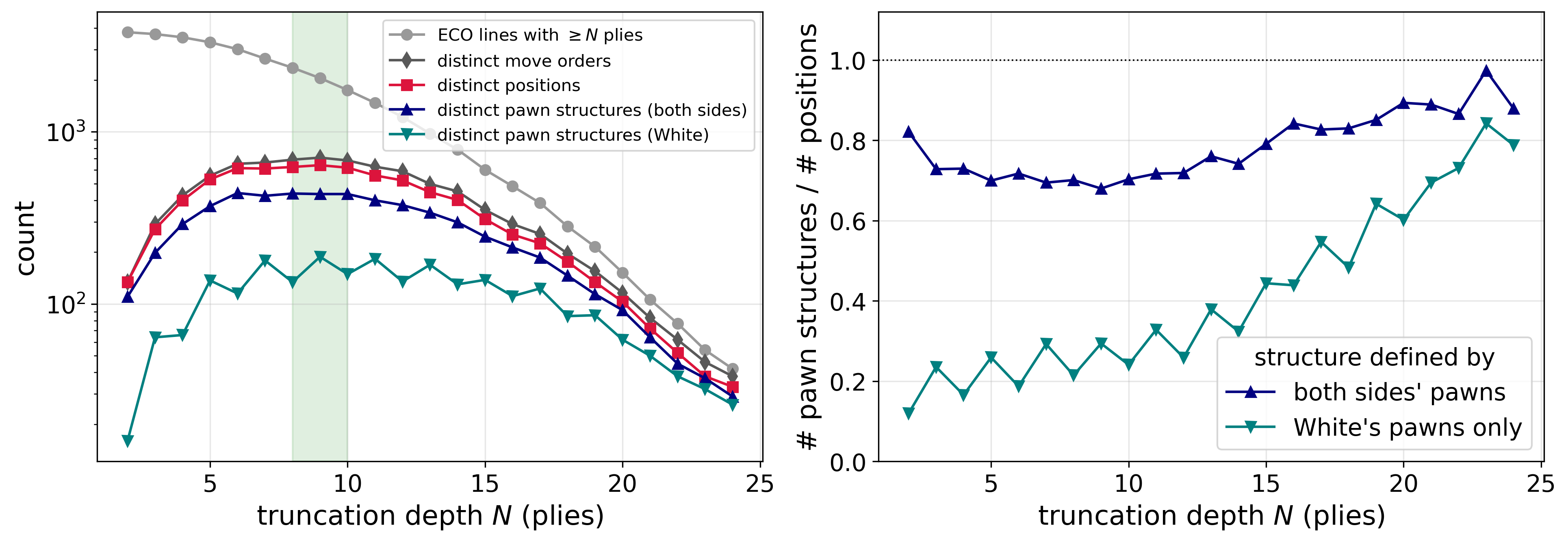}
\caption{
\textbf{From named lines to coarse-grained states.}
\emph{Left:} number of named lines surviving to depth $N$ (number of plies), distinct
truncated move sequences, distinct resulting positions, and distinct
pawn skeletons. The number of distinct positions is maximal around
$N=8$--$10$, which motivates the working depth.
\emph{Right:} ratio between the number $\Sigma(N)$ of distinct pawn
skeletons and the number $\Pi(N)$ of distinct positions. For the joint White--Black pawn skeleton, $\Sigma/\Pi\simeq0.70$ throughout the main working range, showing that the pawn configuration provides a substantial
but incomplete coarse graining of opening space.
}
\label{fig:depth}
\end{figure*}

The choice is based on the empirical findings shown in Fig.~\ref{fig:depth}. At small $N$, many named
lines have not yet separated, whereas at large $N$, many catalogue lines no
longer survive because their definitions are shorter than the chosen
depth. The number of distinct positions is therefore nonmonotonic in
$N$ and reaches a broad maximum around $N\approx8-10$.

At $N=10$, 1749 of the original 3810 named lines survive, and they reduce as
\begin{equation}
\nonumber
1749\ {\rm names}
\longrightarrow
682\ {\rm move\ sequences}
\longrightarrow
619\ {\rm positions}.
\label{eq:collapse}
\end{equation}

Most of the reduction occurs between names and move sequences: many
catalogue names share the same first ten plies and separate only later. The remaining reduction is due to different move
sequences that can sometimes reach exactly the same board position. In most of such cases, the same moves have merely been played in a different order.
Indeed, if move order is ignored, the 682 sequences reduce to 621
distinct collections of moves, comparable with the 619 final
positions. Note that the oscillations in the number of White pawn structures observed in Fig.~\ref{fig:depth} arise from the alternation of moves: White pawn moves introduce new structures at odd plies, whereas at even plies the White pawn structure generally remains unchanged. This is combined with the progressive loss of shorter opening lines from the catalog, and produces the observed alternating increases and decreases.

Positions, rather than names, are therefore the elementary states in
what follows, whereas traditional opening names are retained as metadata.

In order to identify uniquely positions, we use the coordinates system as follows. Let $\mathcal S$ be the set of 64 squares, where a square $s$ has file
$f(s)\in\{0,\ldots,7\}$ and absolute rank
$r(s)\in\{0,\ldots,7\}$. For a side $c\in\{W,B\}$ we define the forward rank
\begin{equation}
\rho_c(s)=
\begin{cases}
r(s), & c=W,\\
7-r(s), & c=B,
\end{cases}
\label{eq:rho}
\end{equation}
so that the two colors are treated symmetrically. The pawn home rank is
$\rho_c=1$ for both sides.

\begin{figure*}[!t]
    \centering
    \includegraphics[width=\textwidth]{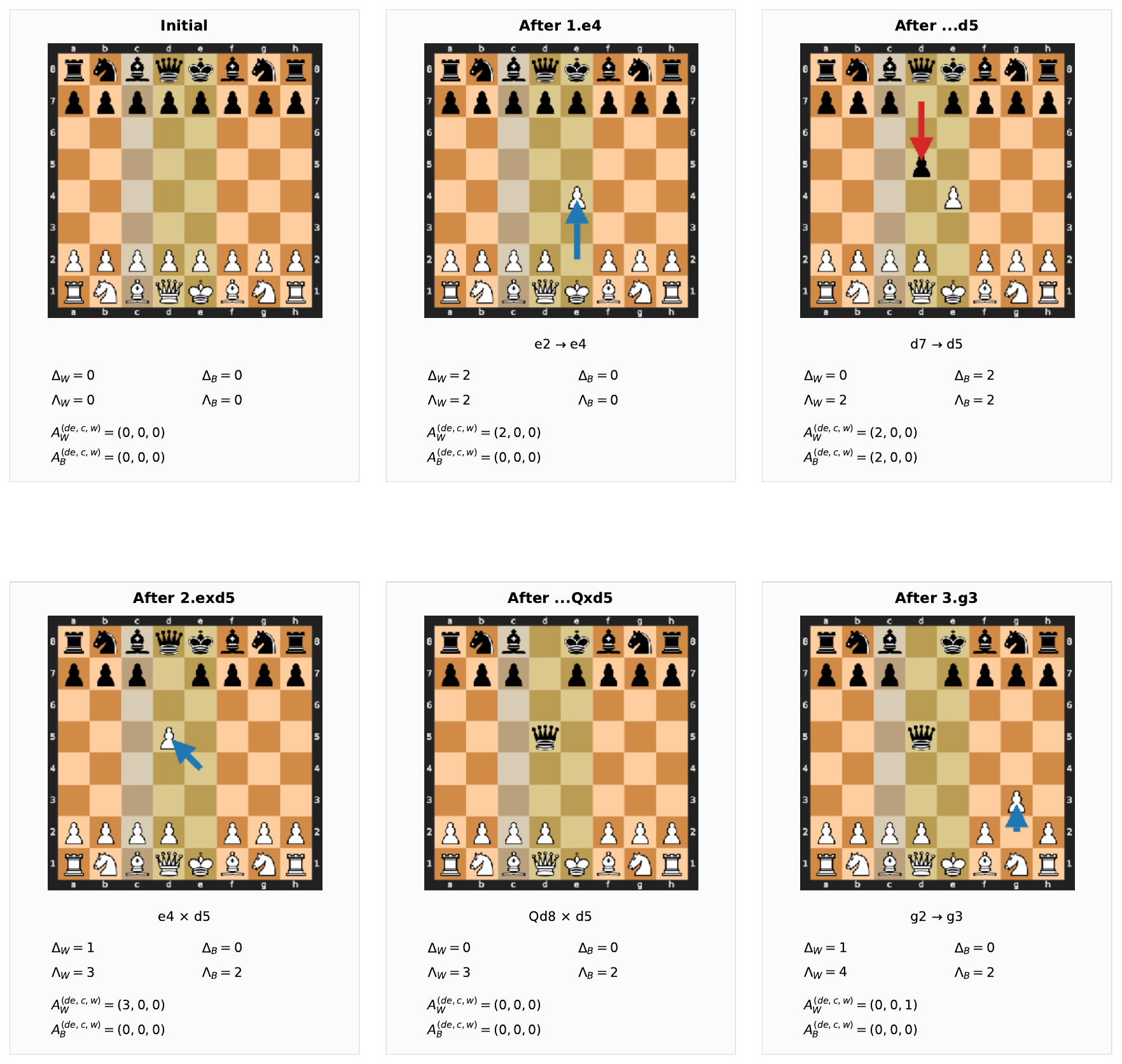}
    \caption{\textbf{Cumulative pawn commitment and spatial allocation of surviving pawn advances}. Six consecutive positions from the opening sequence \texttt{1.e4 d5 2.exd5 Qxd5 3.g3} illustrate the distinction between the historical and configuration-dependent observables introduced in the text. The increments $\Delta_W$ and $\Delta_B$ measure pawn advancement at each move, while $\Lambda_W$ and $\Lambda_B$ accumulate these contributions, including those of subsequently captured pawns. The vectors $\mathbf{A}_W$ and $\mathbf{A}_B$ quantify the advancement of surviving pawns across the central files (d,e), the c-file, and the remaining wing (a,b,f,g,h) files, respectively. Shaded columns identify the central and c-file regions. Unlike $\Lambda_c$, the spatial observables can decrease following captures, revealing the distinction between irreversible historical commitment and the current pawn structure.}
    \label{fig:sketch}
\end{figure*}

\section{Pawns: the slow structure}
\label{sec:slow}

Pawns provide the clearest slow degrees of freedom of a chess position. Unlike pieces, which can often be redeployed in a few moves, pawn advances are irreversible and progressively constrain the geometry of the board, and retain a memory of the opening trajectory while simultaneously defining its current structural skeleton. We characterize these two aspects separately: the move history measures the accumulated pawn commitment, whereas the board position describes the morphology of the surviving pawn structure.

\subsection{Irreversible commitment}

For each side $c$, we define the cumulative pawn commitment as the total number of ranks advanced by its pawns up to the position considered:
\begin{equation}
\Lambda_c =
\sum_{\substack{\text{pawn moves }m\\
                \text{by }c}}
\Delta\rho_{c,m},
\label{eq:Lambda}
\end{equation}
where $\Delta\rho_{c,m}=\rho_c(d(m))-\rho_c(o(m))$ and 
$d(m)$ and $o(m)$ are the destination and origin squares. Thus, a normal pawn move contributes $1$, while an initial two-square move contributes $2$. The quantity $\Lambda_c$ therefore measures the total irreversible pawn advance accumulated during the opening, including moves made by pawns that may subsequently have been captured. Because Eq.~(\ref{eq:Lambda}) is evaluated from
the move history, a contribution remains even if that pawn is later
captured. $\Lambda_c$ is therefore nondecreasing with depth and records
accumulated irreversible structural decisions.

Note that simply counting pawn moves would be less informative, because at fixed $N$, each
side has the same move budget, so such a count would mainly tell us how many moves were spent on pawns rather than on other pieces.
Equation~(\ref{eq:Lambda}) instead measures the total distance advanced
and keeps the history of pawns that have disappeared from the board.

\subsection{Where the pawns go}

In order to understand where the surviving pawn advancement is located in the current position, we need to introduce additional indicators.

For each pawn $p$ still on the board, let $a_p$ be the number of ranks it
has advanced from its initial rank. Thus, a pawn on its home rank has
$a_p=0$, a pawn one rank forward has $a_p=1$, and so on. We divide the
board into three groups of files: the central $d$ and $e$ files, the $c$
file, and the remaining wing files. We then define
\begin{align}
A^{de}_c
&=
\sum_{\substack{p\text{ pawn of }c\\
                 \mathrm{file}(p)\in\{d,e\}}}
a_p,
\\
A^{c}_c
&=
\sum_{\substack{p\text{ pawn of }c\\
                 \mathrm{file}(p)=c}}
a_p,
\\
A^{w}_c
&=
\sum_{\substack{p\text{ pawn of }c\\
                 \mathrm{file}(p)\in\{a,b,f,g,h\}}}
a_p.
\label{eq:allocation}
\end{align}

The three quantities measure how the surviving pawn advancement is
distributed across the board: $A^{de}_c$ measures advancement in the
center, $A^{c}_c$ on the $c$ file, and $A^{w}_c$ on the wings. For example, a White pawn on \texttt{d4} contributes $2$ to
$A^{de}_W$, whereas a White pawn on \texttt{d2} contributes $0$. These observables differ fundamentally from $\Lambda_c$ (see Fig.~\ref{fig:sketch}). While
$\Lambda_c$ retains the full history of pawn advances, the $A$'s depend
only on the pawns present in the current position: once a pawn is
captured, its contribution disappears. They therefore describe the
spatial allocation of the surviving pawn structure rather than its
cumulative history.

The $c$ file is treated separately because its behavior in the opening
catalogue is empirically distinct from both direct central occupation and
wing expansion.

\subsection{Pawn contact: blockade and tension}

Some properties of a position depend on the relative geometry of the two
pawn skeletons. We distinguish two kinds of direct contact: \emph{blocked pairs} ($B$), in which
opposing pawns face one another directly, from \emph{levers} ($L$), in
which opposing pawns are diagonally adjacent. The blocked pairs form a locally frozen structure, while a lever represents an unresolved tension that can
potentially be released through captures.

We use the total number of contacts,
\begin{equation}
\mathcal C=B+L,
\label{eq:contact}
\end{equation}
as a coarse indicator of pawn contact and define
\begin{equation}
\begin{cases}
\mathcal C=0, & \text{open},\\
\mathcal C=1, & \text{semi-open},\\
\mathcal C\geq 2, & \text{closed}.
\end{cases}
\label{eq:regime}
\end{equation}
These labels refer to pawn contact rather than to openness in the full
chess sense. At $N=10$, the 619 distinct positions comprise 309 open,
177 semi-open, and 133 closed positions according to this definition.

\begin{figure}[t]
\centering
\includegraphics[width=\columnwidth]{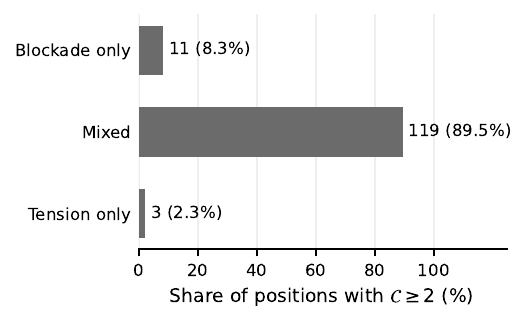}
\caption{\textbf{Blockade and tension in high-contact positions.}
Composition of the 133 distinct positions at $N=10$ with
$\mathcal C=B+L\geq2$: blockade only ($B>0,L=0$), mixed
($B>0,L>0$), and tension only ($B=0,L>0$).
Bars show the percentage of positions; annotations give counts and
percentages.}
\label{fig:contact_types}
\end{figure}

The distinction between blockade and tension is important even when
$\mathcal C$ is fixed. For $\mathcal C>0$, the lever fraction
$\eta=L/\mathcal C$ measures their relative contribution. As shown
in Fig.~\ref{fig:contact_types}, among the 133 positions with
$\mathcal C\geq2$, 119 ($89.5\%$) contain both blocked pairs and
levers, while only 11 contain blockade alone and three contain
levers alone. Moreover, 109 ($82.0\%$) have $B=L$
(i.e., $\eta=1/2$). Thus the high-contact class predominantly
consists of mixed structures, with static constraints and
potentially resolvable tension coexisting. We therefore retain
$B$ and $L$ separately in the morphological description.

\subsection{Pawn structure as a coarse graining}
\label{sec:coarse}

The slow--fast picture is useful only if the pawn structure is
informative but incomplete, and the data show precisely this intermediate
behavior. Let $\Pi(N)$ be the number of distinct full positions reached at depth
$N$, and let $\Sigma(N)$ be the number of distinct joint White--Black
pawn skeletons among those positions. At $N=10$, we obtain $\Pi=619$, and $\Sigma=435$, so that
\begin{equation}
\frac{\Sigma}{\Pi}
=
\frac{435}{619}
\simeq0.70.
\label{eq:coarse_ratio}
\end{equation}
(the $\Sigma=435$ distinct joint pawn configurations are defined by the exact locations of White and Black pawns and these configurations should not be confused with the much smaller number of conventional pawn-structure families used in chess theory, which group together strategically similar arrangements). Equivalently, one joint pawn skeleton represents on average
$\Pi/\Sigma\simeq1.42$ full positions. This ratio is rather stable
over the useful depth range: for $N=6,8,10,$ and $12$ it is approximately
$0.72$, $0.70$, $0.70$, and $0.72$, respectively
(see Fig.~\ref{fig:depth}). Therefore, projecting the full position onto the pawn skeleton merges roughly $30\%$ of the distinct states while retaining the rest. In the elementary statistical-physics sense, the pawn skeleton
is a genuine coarse-grained variable: several different piece configurations can live on the same slower structural background, and the residual multiplicity within a pawn skeleton is precisely what the
fast sector will describe.

\subsection{A network of pawn skeletons}
\label{sec:skelnet}

The previous analysis treats a pawn skeleton as a coarse-grained state, and we can also ask how the realized skeletons are related to one another.
For this purpose we construct a network for White at fixed $N=10$. There are 149 distinct White pawn skeletons among the 619 positions, so
one White skeleton is shared on average by 4.15 positions. The most common one occurs in 42 positions.


Consider two White pawn skeletons, $A$ and $B$, where each skeleton is
the set of squares occupied by White pawns. We draw a directed link
\begin{equation}
A\longrightarrow B
\end{equation}
when the structure $A$ can be transformed into $B$ by a single
irreversible pawn event, according to the two following possibilities.

\paragraph{Pawn move.}
The two skeletons contain the same number of pawns, and all pawns occupy
the same squares except one. In $A$ this pawn is on a square $s$, while
in $B$ it is on a square $t$. We add a link $A\to B$ if $s\to t$ has
the geometry of a possible White pawn move: one square forward, two
squares forward from the initial rank, or one square diagonally forward (in the case of capture).
Schematically,
\begin{equation}
A=\{\ldots,s,\ldots\}
\quad\longrightarrow\quad
B=\{\ldots,t,\ldots\}.
\label{eq:pawn_move_link}
\end{equation}

\paragraph{Pawn loss.}
The skeleton $B$ contains exactly one pawn fewer than $A$, while all
remaining pawns occupy the same squares. This corresponds to one White
pawn being captured by Black:
\begin{equation}
A=\{\ldots,s,\ldots\}
\quad\longrightarrow\quad
B=\{\ldots\}.
\label{eq:pawn_capture_link}
\end{equation}

Thus a link $A\to B$ means that the two structures differ by only one
elementary pawn event: either one pawn has advanced, or one pawn has
disappeared after being captured. The links are directed because these
changes cannot be reversed by pawn dynamics: pawns cannot move backward,
and captured pawns cannot return.

In the following subsections, we characterize the structure of the resulting network by means of complex network analysis.

\subsubsection{A giant connected component}

\begin{figure*}[t]
\centering
\includegraphics[width=\textwidth]{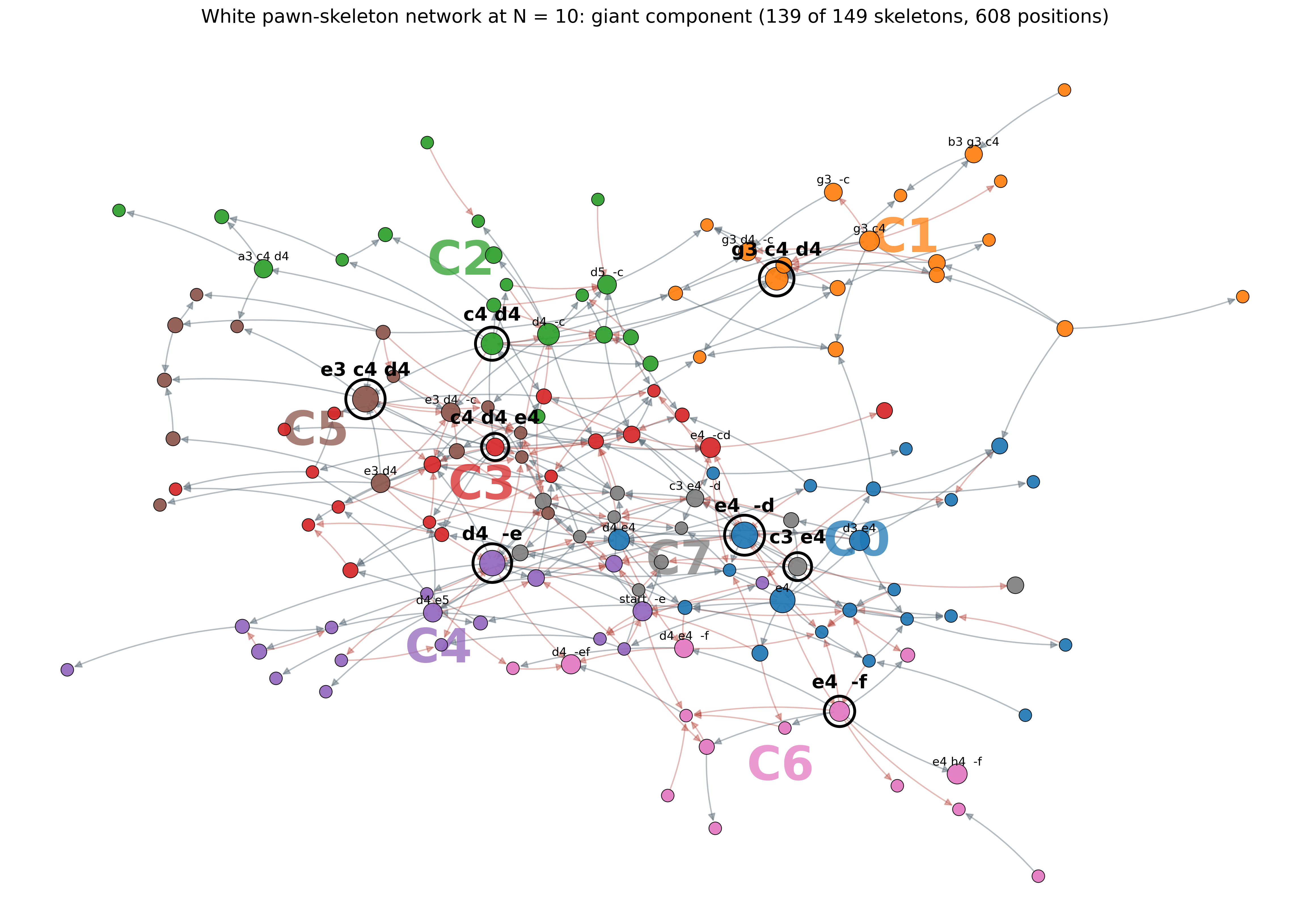}
\caption{
\textbf{Network of White pawn structures at $N=10$.}
The giant weakly connected component contains $139$ of the $149$
distinct White pawn skeletons and represents $608$ of the $619$
positions ($98\%$). Nodes are pawn skeletons, with node size proportional
to the number of positions carrying that structure. The network contains
$335$ directed links, corresponding to single pawn moves or pawn losses
through capture; for community detection, link direction is ignored.
Colors show the eight communities identified by the Leiden algorithm,
with sizes ranging from $12$ to $22$ skeletons. The observed modularity,
$Q=0.569$, is substantially larger than for degree-preserving randomized
networks, $Q_{\rm null}=0.444\pm0.007$ ($z\simeq17.5$), showing that the
community organization is significantly stronger than expected from the
degree sequence alone. The ringed node in each community is its medoid,
i.e.\ the skeleton with the smallest mean graph distance to the other
members of that community. The ten isolated skeletons are not shown.
}
\label{fig:skeleton_network}
\end{figure*}

The network contains $149$ distinct White pawn skeletons connected by $335$
directed links, of which $199$ ($59\%$) correspond to pawn moves and $136$
($41\%$) to pawn captures (Fig.~\ref{fig:skeleton_network}). The mean
in-degree and out-degree are both $335/149\simeq 2.25$, with $11$ weakly
connected components. One of them is a giant component containing $139$
skeletons, while the other $10$ components consist of isolated single nodes. 
Since these isolated nodes carry no links, all $335$ links belong to the giant
component, corresponding to a mean in- and out-degree of $335/139\simeq2.41$
within the connected backbone.

The giant component therefore contains $139/149\simeq93\%$ of all White
pawn skeletons and represents $608$ of the $619$ positions, or $98\%$ of the
catalogue at $N=10$. Thus, despite the diversity of named openings, almost all
positions belong to a single connected system of pawn structures linked by
elementary pawn moves or captures. The ten isolated skeletons account for the
remaining $11$ positions.

These isolated states correspond to early pawn configurations. Several occur in gambit lines, such as the Portuguese Gambit, Sicilian Wing
Gambit, and Evans Gambit Declined, while others occur in openings such as the
Alekhine Defense, Hippopotamus, Grob, Polish Opening, Modern Defense, and
Queen's Gambit Declined. These openings are mostly distinguished by the absence from the catalogue of any pawn skeleton connected to them by a single elementary pawn event.

\subsubsection{Mesoscale organization}

A distinctive feature of complex network is their nontrivial organization into clusters or modules \cite{fortunato2010community}. One way to unravel such an organization is to ask whether the giant component contains groups of pawn
structures that are more densely connected to one another than to the
rest of the network. We first ignore link direction and treat an elementary
pawn transformation simply as a neighborhood relation in pawn-structure
space. We then apply the Leiden community-detection algorithm to this
undirected network \cite{Blondel2008,Traag2019}. The algorithm directly
optimizes the Newman--Girvan modularity $Q$ \cite{Newman2006} and determines
the number of communities without requiring it to be specified in advance. 
Nevertheless, the effect of retaining the original direction of the links is tested below.

We identify $K=8$ communities, with sizes 22, 19, 19, 19, 18, 16, 14, 12, and modularity $Q_{\textrm{obs}}=0.569$. Since a positive modularity alone does not demonstrate significant community structure, we compare the observed network against an ensemble of degree-preserving null networks obtained by random edge swaps. The Leiden algorithm is
run independently on each randomized network, so that both the number of
communities and their composition are free to vary: over $1000$ null realizations, we obtain
$Q_{\rm null}=0.444\pm0.007$, where the uncertainty denotes the standard deviation across realizations. The observed modularity $Q_{\textrm{obs}}$ therefore lies about $17.5$ null standard deviations above the randomized expectation, suggesting the existence of a genuine nontrivial modular organization.The null networks contain on average $8.37$ communities
(range $5$--$12$), comparable to the $K=8$ communities
observed in the empirical network. The main difference
therefore lies not in the number of communities, but
in their significantly stronger modular organization
in the empirical pawn-skeleton network.
\begin{table*}[t]
\centering
\small
\setlength{\tabcolsep}{7pt}
\begin{tabular}{clrrrrl}
\toprule
$C$ & medoid & skel. & pos. & med. pos. & mass & origin \\
\midrule
0 & \texttt{e4 $-$d}
  & 22 & 131 & 42
  & \texttt{e4 $-$d}
  & \texttt{e4}
\\

1 & \texttt{g3 c4 d4}
  & 19 & 85 & 22
  & \texttt{g3 c4 d4}
  & \texttt{g3 c4}
\\

2 & \texttt{c4 d4}
  & 19 & 84 & 18
  & \texttt{d4 $-$c}
  & \texttt{c3 d4}
\\

3 & \texttt{c4 d4 e4}
  & 19 & 55 & 7
  & \texttt{e4 $-$cd}
  & \texttt{c4 d4 e4}
\\

4 & \texttt{d4 $-$e}
  & 18 & 82 & 36
  & \texttt{d4 $-$e}
  & \texttt{e5}
\\

5 & \texttt{e3 c4 d4}
  & 16 & 77 & 39
  & \texttt{e3 c4 d4}
  & \texttt{e3 d4}
\\

6 & \texttt{e4 $-$f}
  & 14 & 56 & 12
  & \texttt{e4 h4 $-$f}
  & \texttt{e4 $-$f}
\\

7 & \texttt{c3 e4}
  & 12 & 38 & 8
  & \texttt{c3 e4}
  & \texttt{c3 e4}
\\
\bottomrule
\end{tabular}

\caption{
\textbf{Communities of the White pawn-skeleton network at $N=10$.}
The Leiden algorithm identifies eight communities within the giant
component, listed here in decreasing order of the number of skeletons.
$C$ is an arbitrary community label; ``skel.'' is the number of distinct
White pawn skeletons in the community, and ``pos.'' is the number of
opening positions associated with those skeletons. The ``medoid'' is the
skeleton with the smallest mean graph distance to the other members of
the same community, and ``med. pos.'' gives the number of opening
positions having exactly that skeleton. ``Mass'' denotes the most
frequently occurring skeleton in the community, whereas ``origin'' is an
upstream skeleton identified using the original direction of the network
links. Skeleton labels describe deviations from the initial White pawn
configuration. A square such as \texttt{e4} indicates a White pawn on
that square, while \texttt{$-d$} means that White has no pawn remaining
on the $d$ file; pawns still on their initial squares are omitted. Thus
\texttt{e4 $-$d}, for example, denotes a skeleton with a White pawn on
$e4$ and no White pawn anywhere on the $d$ file.
}
\label{tab:communities}
\end{table*}

\subsubsection{Robustness to direction and opening depth}
\label{sec:community_robustness}

Since our aim here is to identify morphological regions rather than
temporal pathways through opening space, the undirected representation used above is the natural one for the main analysis.

The original network nevertheless carries a genuine direction, since pawn
advances and captures are irreversible. In fact, all $335$ links in the
$N=10$ giant component are one-way, with no reciprocal pairs, and the
resulting directed graph is acyclic. Direction therefore contains additional
information about the allowed evolution of pawn structures, and provides a
natural test of whether the morphological organization found above is an
artifact of discarding irreversibility.

We repeated the Leiden analysis on the directed graph using directed
modularity, finding $K_{\rm dir}=7$ communities and
$Q_{\rm dir}=0.575$. This value should not be ranked directly against the
undirected modularity $Q=0.569$, since the two quantities are assessed with respect to different baseline expectation. We compare the two partitions using two complementary measures.
The normalized mutual information (NMI) \cite{Danon2005}
quantifies how much information about one partition is retained
in the other, while the adjusted Rand index (ARI)
\cite{HubertArabie1985} measures agreement in the assignment
of pairs of nodes to communities, corrected for chance.
These quantities are both equal unity for identical partitions, but respond differently to the splitting or merging of communities.
We find $\mathrm{NMI}=0.852$ and $\mathrm{ARI}=0.730$,
showing that the community structure is largely preserved
when edge directions are taken into account.

For each community, we define its medoid as the pawn skeleton that minimizes
the mean graph distance to all other skeletons in that community. Six of the
eight medoids identified in the undirected network are also recovered in the
directed analysis, and several communities are unchanged or nearly unchanged.

\begin{figure*}[t]
\centering
\includegraphics[width=0.47\textwidth]{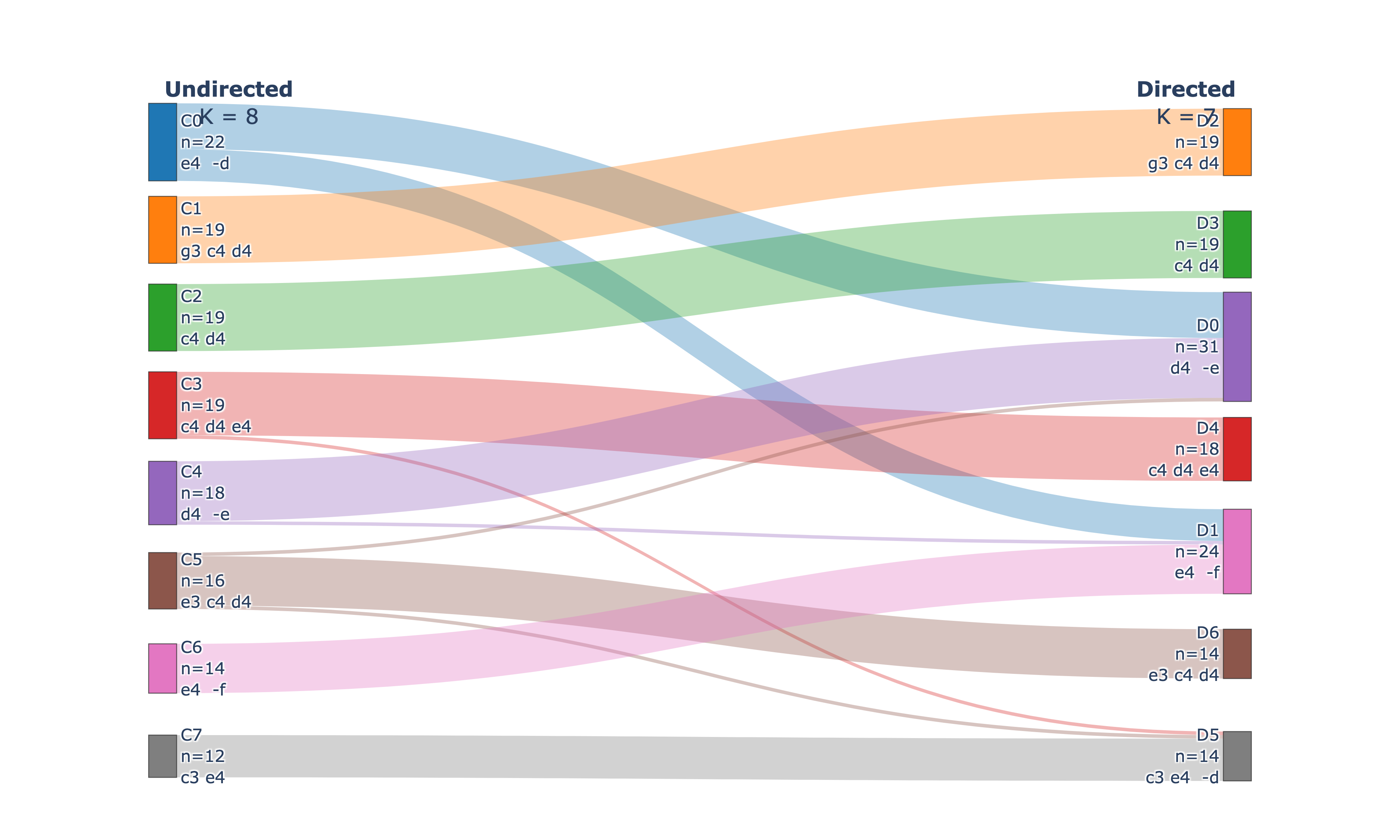}
\includegraphics[width=0.47\textwidth]{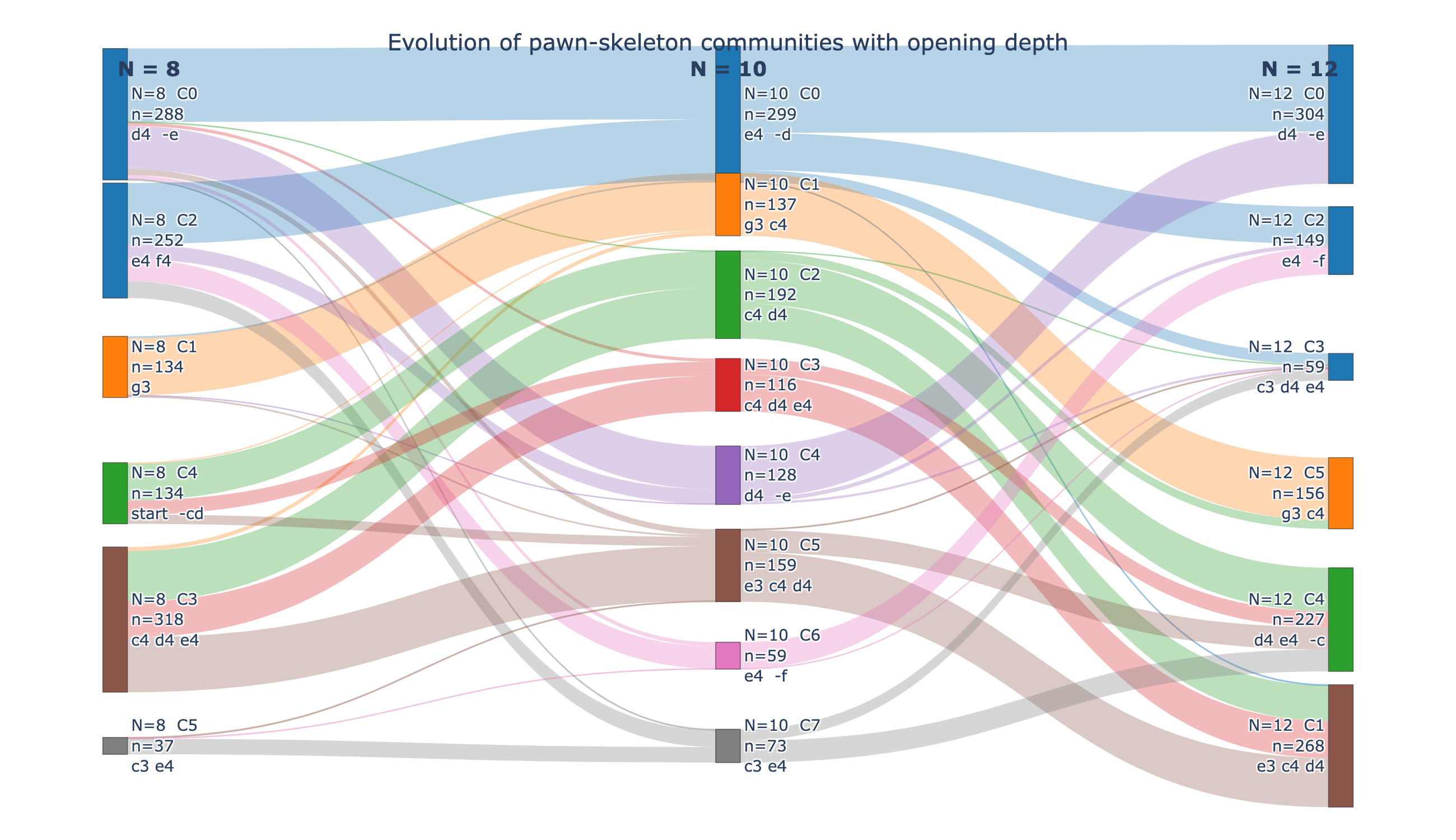}
\caption{
\textbf{Robustness and evolution of the pawn-skeleton communities.}
(a) Comparison of the undirected and directed Leiden partitions of the
$N=10$ giant component. Ribbon widths represent the number of pawn
skeletons shared between communities. Retaining link direction changes the
number of communities from eight to seven, while the two partitions retain
substantial agreement: the normalized mutual information is
${\rm NMI}=0.852$ and the adjusted Rand index is ${\rm ARI}=0.730$. Six of the eight
undirected medoids are also recovered as directed medoids.
(b) Evolution of community membership with opening depth for the same
catalogue lines. The diagram follows the $1163$ openings that survive to
$N=12$ and belong to the giant component at all three depths
$N=8,10,$ and $12$; ribbon widths represent numbers of openings.
The corresponding undirected partitions have
$(K,Q)=(6,0.559)$, $(8,0.569)$, and $(6,0.588)$, respectively.
The flows show that modular organization persists with opening depth,
although individual communities split and merge as the pawn structure
develops.
}
\label{fig:community_robustness}
\end{figure*}

The main difference concerns one of the largest undirected communities, whose members
are redistributed between two directed groups. Retaining link direction
therefore modifies some community boundaries, but leaves the broad
organization of pawn-structure space largely unchanged
[Fig.~\ref{fig:community_robustness}(a)]. This result thus supports the use in the following of the undirected graph as the main representation of morphological proximity.

We next examine how this organization evolves with opening depth. Repeating
the undirected analysis at $N=8,10,$ and $12$ gives for the number of communities and the modularity
\begin{equation}
\nonumber
(K,Q)=(6,0.559),\,(8,0.569),\, (6,0.588),
\end{equation}
respectively. The modularity therefore remains of comparable magnitude over
this range, while the number and boundaries of the detected communities
change with depth. The number of distinct White pawn skeletons in the full
network also changes non-monotonically,
\begin{equation}
\nonumber
134 \longrightarrow 149 \longrightarrow 135,
\end{equation}
for $N=8,10,$ and $12$, respectively.

A direct comparison based only on identical skeletons at different depths is
too restrictive, since the pawn skeleton of a given opening itself evolves
with $N$, and we therefore compare the community assignments of the same catalogue lines. Among the $1219$ lines that survive to $N=12$, $1163$
($95.4\%$) belong to the giant component at all three depths. Following these
same openings from $N=8$ to $N=10$ and $N=12$ gives
\begin{equation}
\begin{array}{c|ccc}
(N_1,N_2) & (8,10) & (10,12) & (8,12) \\
\hline
{\rm NMI} & 0.529 & 0.579 & 0.585 \\
{\rm ARI} & 0.350 & 0.393 & 0.527
\end{array}
\label{eq:depth_partition_stability}
\end{equation}
indicating partial, rather than exact, persistence of community membership.

The alluvial representation in
Fig.~\ref{fig:community_robustness}(b) shows that this evolution is more complex than a simple loss of community structure. Broad families present at $N=8$ differentiate
into a finer set of structural groups by $N=10$, while several of these
groups subsequently merge into broader regions at $N=12$. Some lineages
remain highly coherent throughout this evolution, whereas others branch
substantially between successive depths. Within the range studied, $N=10$
therefore appears as a particularly differentiated stage of pawn-structure
space: both the number of detected communities and the number of distinct
skeletons reach their largest values there.

We do not interpret the eight $N=10$ communities as universal or
depth-independent classes. Rather, they provide one resolution of a broader,
evolving morphology in which pawn structures first differentiate as
irreversible commitments accumulate and then partly consolidate as distinct
opening histories converge toward common structural regions.

\subsubsection{Relation to classical pawn structures}
\label{sec:classical_pawn_structures}

Pawn structure plays a central role in classical chess strategy. Because
pawn moves are largely irreversible, pawn configurations tend to persist
and constrain the plans available to both players. This idea already appears
in the foundations of modern positional theory. In \emph{The Modern Chess
Instructor}, the first World Chess Champion W. Steinitz (1836-1900) emphasized pawn structure, pawn weaknesses, and weak
squares as lasting positional features, in contrast with more transient
advantages such as development or initiative \cite{Steinitz1889}.
Chess theory subsequently organized many positions around recurring pawn
structures and the strategic ideas associated with them
\cite{Soltis1995,FloresRios2015}. This viewpoint fits naturally with the
slow--fast description developed here: pawns define a slowly evolving
structural background, while pieces can be reorganized more rapidly within
the geometry it imposes.

The eight communities detected in the White pawn-skeleton network at $N=10$
are summarized in Table~\ref{tab:communities}. 
These communities emerge from the pawn-skeleton network alone,
without using opening names or predefined classifications.
Nevertheless, their representative skeletons correspond to
familiar opening patterns. We represent each community by
its medoid, defined as the skeleton with the smallest average
graph distance to the other members of the community.
This choice reflects the geometry of the network rather
than the frequency of opening names in the catalogue. For
comparison, Table~\ref{tab:communities} also reports the most frequently
represented skeleton (`mass'') and an upstream skeleton (`origin'')
defined from the original direction of the links.

In fact, several of the resulting communities have straightforward chess
interpretations. The community centered on \texttt{g3 c4 d4} contains
fianchetto structures characteristic of many English and Catalan systems.
Those centered on \texttt{c4 d4} and \texttt{e3 c4 d4} belong to the broad
family of queen's-pawn structures from which Queen's-Gambit, Slav, and
related formations develop, while \texttt{c4 d4 e4} corresponds to positions
in which White has established a broad pawn center.

Other communities are organized around an early $e4$ structure, with the largest having medoid \texttt{e4 $-$d}, corresponding to White that retains the pawn on
$e4$ but no longer has a pawn on the $d$ file. The communities centered on
\texttt{e4 $-$f} and \texttt{c3 e4} represent different developments of
this basic structure. Finally, the community centered on \texttt{d4 $-$e}
is dominated by positions in which White retains a $d$ pawn while the
$e$-file pawn has disappeared. The network therefore distinguishes several
recurring ways in which the central pawn structure is established and
transformed during the opening.

However, as mentioned earlier, these correspondences should not be understood as exact identifications with the named pawn formations of chess theory. Our nodes encode only the
White pawn skeleton, whereas classical pawn structures are generally
defined by the relative arrangement of the pawns of both sides. A single
White skeleton may therefore belong to different classical structures
depending on Black's configuration. Moreover, at $N=10$ many mature
middlegame formations, such as the isolated queen's pawn, hanging pawns,
or the Mar\'oczy Bind, need not yet have formed. The detected communities
are therefore better interpreted as “early structural families” from which more specific middlegame structures may develop.

The links of the network also give a precise meaning to the classical idea
that pawn structures transform into one another
\cite{Soltis1995,FloresRios2015}. Two neighboring skeletons differ by a
single elementary pawn event: a pawn advance or the disappearance of a pawn
after capture. The directed network thus maps the local structural
transformations actually realized in opening theory, and goes beyond a list of named pawn structures: it reveals a continuous structural organization underlying the more discrete taxonomy of classical opening theory.

\section{The fast sector: measuring the piece activity}
\label{sec:fast}

The pieces other than pawns -- such as knights, bishops, rooks, and queens -- form the fast sector of an opening position and, unlike pawns, they can often be rerouted or redeployed without altering the underlying pawn structure. Their organization is therefore better described by their current activity on the board than by their history of moves. We characterize this activity through the squares controlled by the pieces.
This representation contains two distinct pieces of information: how much
activity the pieces generate, and where that activity is directed. In the following, the king is excluded because its location and mobility are dominated by king safety and castling,
rather than by the same notion of piece activity.


For each color $c$, let $m_c(s)$ denote the number of non-pawn,
non-king pieces of color $c$ attacking square $s$. The resulting field
$m_c(s)$ gives a spatial representation of piece activity over the board:
several pieces may contribute to the same square, while a single piece
typically contributes to many squares. Accordingly, the total attack mass (at a given position on the board)
\begin{equation}
M_c=\sum_s m_c(s)
\label{eq:M}
\end{equation}
measures the overall amount of activity generated by the pieces. It is
simply the total number of piece--square attacks in the position, with
blocking pieces and the usual chess movement rules (taken into account as
implemented in \texttt{python-chess} \cite{PythonChess}). Thus $M_c$ measures how extensively the pieces project their influence
over the board, but it does not distinguish whether this activity is
directed toward the center, the flanks, or the player's own camp.


In order to capture the spatial organization of this activity, we define the
centralization
\begin{equation}
\kappa_c
=
\frac{
\sum_s G(s)m_c(s)
}{
\sum_s m_c(s)
},
\label{eq:kappa}
\end{equation}
where central squares are given larger weight. Rather than imposing a
sharp distinction between central and noncentral squares, we use the
smooth Gaussian weight
\begin{equation}
G(s)=
\exp\left[
-\frac{
(f(s)-7/2)^2+(r(s)-7/2)^2
}{2\sigma^2}
\right],
\label{eq:G}
\end{equation}
with $\sigma=3/2$. The weight is then maximal at the center of the board and decreases continuously toward its edges.

Because $\kappa_c$ is normalized by the total attack mass, it measures
where piece activity is concentrated rather than how much activity is
present. Two positions may therefore have similar values of $M_c$ but
different values of $\kappa_c$: in one, the pieces may control many
squares away from the center, while in the other their activity may be
strongly focused toward central squares.

The pair $(M_c,\kappa_c)$ consequently separates two aspects of the fast
sector: the \emph{amount} of piece activity and its \emph{spatial
organization}. These quantities provide a compact description of how the
pieces are deployed on top of the more slowly evolving pawn structure.

\section{Opening morphology}
\label{sec:morphology}

Having introduced the measures for characterizing the slow and fast sectors separately, we now ask how
they combine to organize the space of openings. Before doing so, we first
verify that the two sectors carry genuinely complementary information.

\subsection{Slow and fast are distinct}

For White ($c=W$) at $N=10$, the pairwise correlations among
\begin{equation}
{
\Lambda,
A^{de},
A^c,
A^w,
M,
\kappa
}
\end{equation}
remain below $0.45$ in magnitude, with the exception of the pair $(M,\kappa)$, which gives a correlation of about 0.66. This correlation is expected, since developing and activating the pieces
tends to increase both their total attack mass and the fraction of that
activity directed toward the center.

More importantly, the relation between representative slow and fast
observables is weak. For pawn commitment $\Lambda$ and piece
centralization $\kappa$, we find
\begin{align}
\nonumber
\mathrm{corr}(\Lambda,\kappa)&=0.32
\qquad\text{for White},\\
\nonumber
\mathrm{corr}(\Lambda,\kappa)&=0.06
\qquad\text{for Black}.
\label{eq:slowfastcorr}
\end{align}

Thus, strongly committed pawns need not be accompanied by strongly
centralized pieces, and conversely, high piece centralization does not
imply an advanced pawn structure. The two sectors therefore provide
complementary information: the pawns retain the accumulated structural
history of the opening, while the pieces describe how activity is currently
organized within that structure.

This motivates describing openings in a multidimensional space of slow and
fast coordinates. In the simplest such representation, we use one
coordinate from each sector: pawn commitment $\Lambda$ and piece
centralization $\kappa$.

\subsection{The slow--fast plane}

\begin{figure*}[t]
\centering
\includegraphics[width=0.94\textwidth]{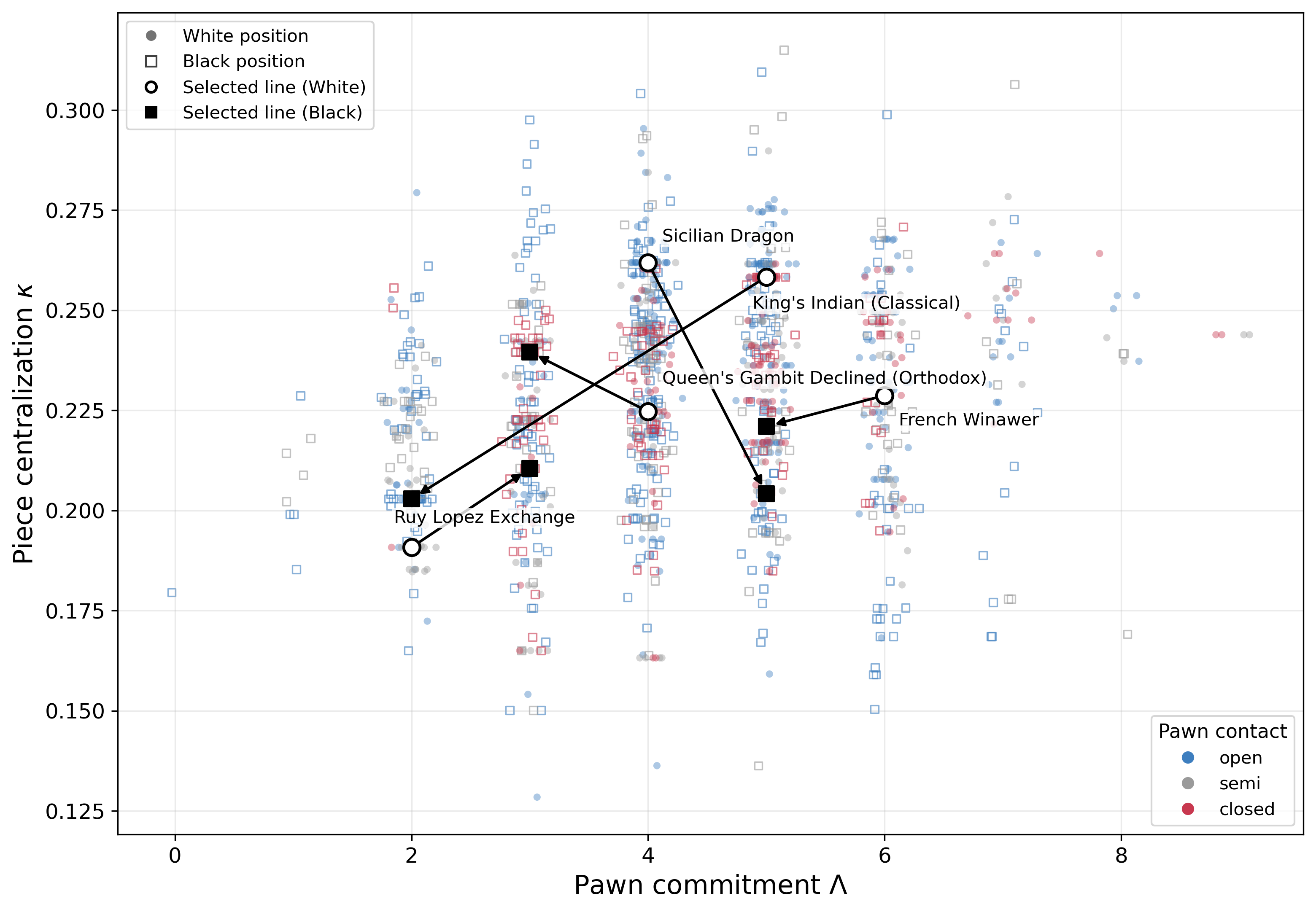}
\caption{
\textbf{A slow--fast map of chess openings.}
Each position at $N=10$ is represented by its pawn
commitment $\Lambda$ and piece centralization $\kappa$,
separately for White (filled circles) and Black
(open squares). Colors indicate the pawn-contact
regime of Eq.~(\ref{eq:regime}).
Larger symbols and arrows highlight five specific
opening variations: Sicilian Dragon, French Winawer,
Ruy Lopez Exchange, Queen's Gambit Declined (Orthodox),
and King's Indian Defense (Classical).
Each arrow connects the White and Black coordinates
of the same position after ten plies.
The highlighted positions are individual examples,
not averages over opening families.
}
\label{fig:headline}
\end{figure*}

Figure~\ref{fig:headline} represents the 619 distinct
positions at $N=10$ in the $(\Lambda,\kappa)$ plane.
Each position contributes two points, one for White and
one for Black. In order to illustrate this representation, we
highlight five familiar opening variations, with arrows
connecting the White and Black coordinates of each
individual position.

Low $\Lambda$ characterizes positions with limited
pawn advancement, whereas high $\Lambda$ indicates
greater irreversible structural commitment.
The range is substantial: some Four Knights and
Italian variations have $\Lambda_W=2$, while
Four Pawns Attack variations of the Alekhine Defense
reach $\Lambda_W=9$.

Piece centralization provides complementary information.
Positions with similar pawn commitment can display
very different degrees of piece centralization,
showing that these two observables capture distinct
aspects of opening organization. The highlighted
examples also illustrate how White and Black can
occupy different regions of this morphological space,
even within the same opening position.


\subsection{White--Black asymmetry}
\label{sec:firstmove}

At $N=10$, White and Black have each made five moves,
allowing their structural observables to be compared
at equal move counts. Averaging over the 619 distinct
positions gives
\begin{align}
\langle\Lambda_W\rangle &= 4.593,
& \langle\Lambda_B\rangle &= 4.021,\\
\langle M_W\rangle &= 36.942,
& \langle M_B\rangle &= 35.178,\\
\langle\kappa_W\rangle &= 0.233,
& \langle\kappa_B\rangle &= 0.226.
\end{align}
White therefore shows somewhat greater pawn commitment,
piece activity, and centralization on average.
The spatial distribution of pawn advancement also differs:
\begin{align}
\langle A_W^{de}\rangle-\langle A_B^{de}\rangle &= +0.578,\\
\langle A_W^{c}\rangle-\langle A_B^{c}\rangle &= +0.181,\\
\langle A_W^{w}\rangle-\langle A_B^{w}\rangle &= -0.405.
\end{align}
Thus White's surviving pawns are, on average, advanced
farther in the central files, while Black's are advanced
farther on the wings.

\begin{figure*}[t]
\centering
\includegraphics[width=0.96\textwidth]{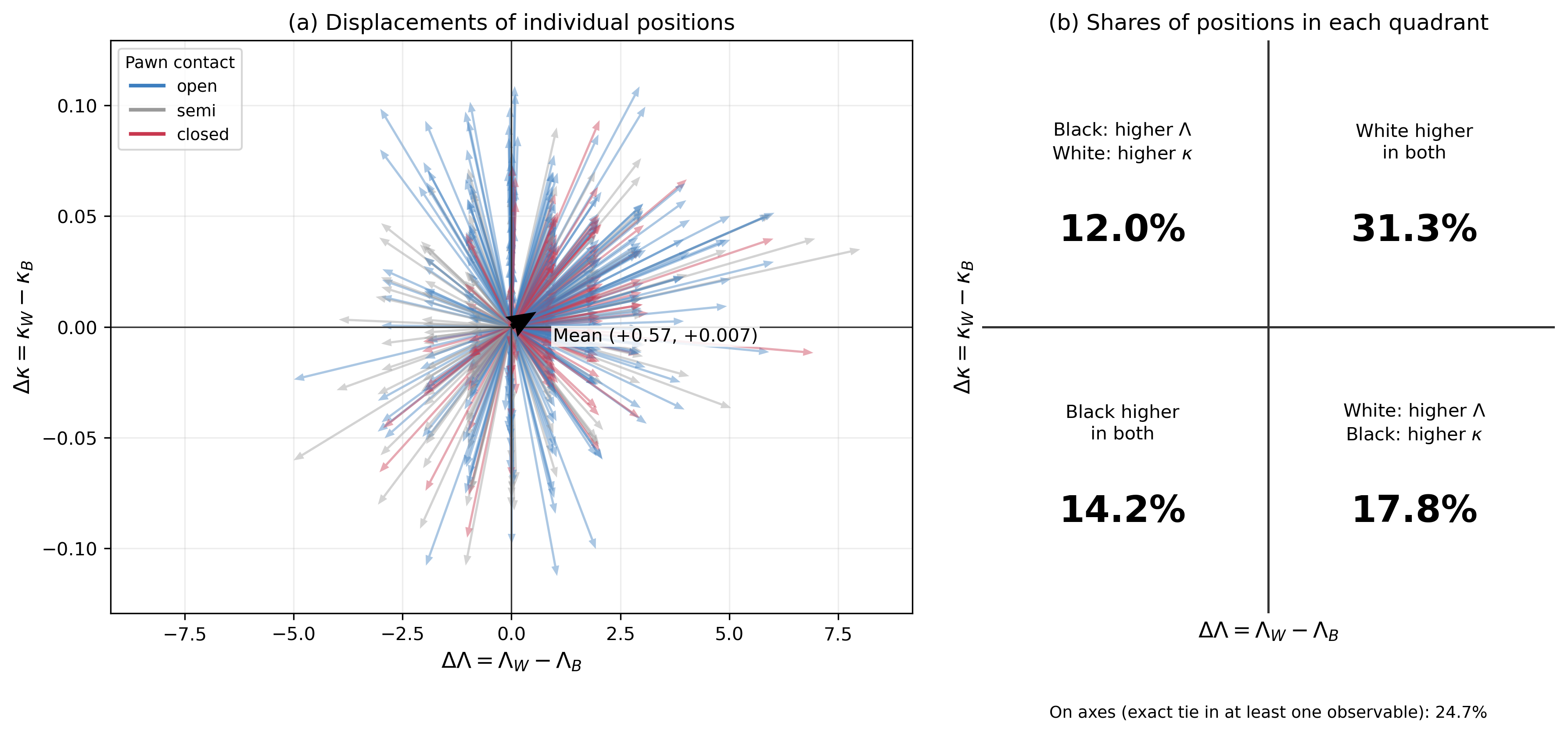}
\caption{
\textbf{White--Black asymmetry in opening morphology.}
(a) Displacement vectors
$\bm v=(\Lambda_W-\Lambda_B,\kappa_W-\kappa_B)$
for the 619 distinct positions at $N=10$,
drawn from a common origin.
Colors indicate the pawn-contact regime
of Eq.~(\ref{eq:regime}); the black arrow
represents the ensemble mean.
A small horizontal jitter separates overlapping arrows.
(b) Percentage of positions in each quadrant of
the same plane (all percentages are
relative to the 619 positions). The upper-right quadrant corresponds
to larger commitment and centralization for White,
and the lower-left to larger values for Black.
The other two quadrants correspond to opposite signs
of the two differences. Positions lying on either axis are excluded from the four quadrants and reported separately. There are
153 such positions ($24.7\%$ of the total), almost
all due to equal pawn commitment ($\Lambda_W=\Lambda_B$).
Only six positions have exactly equal computed
centralization ($\kappa_W=\kappa_B$).
}
\label{fig:vectors}
\end{figure*}

In order to determine whether these differences hold for
individual positions, we define the displacement
\begin{equation}
\bm v =
(\Lambda_W-\Lambda_B,\,
 \kappa_W-\kappa_B),
\label{eq:v}
\end{equation}
with ensemble mean
$\langle\bm v\rangle\simeq(0.57,0.007)$.
Figure~\ref{fig:vectors}(a) shows the individual
displacements from a common origin, together with
their mean.

The distribution is broad compared with its mean.
White has higher pawn commitment and centralization
in $31.3\%$ of positions, whereas Black has higher
values of both in $14.2\%$. The remaining positions
show opposite signs or equality in at least one
observable. Hence the average White--Black asymmetry
is small relative to the variation across individual
positions. 

\section{From morphology to opening choice}
\label{sec:choice}

The physical observables are useful for analysis, but raw values such as
$\Lambda=4.6$ are not immediately meaningful to a player. We therefore
convert each quantity to its percentile within the $N=10$ catalogue.
All coordinates then lie on the same scale from 0 to 100.

\subsection{Scorecards}

In order to quantify the asymmetry between the two pawn structures,
we reflect Black's pawns onto White's side of the board
(preserving the file and reversing the rank) and compare
the two configurations. We define the asymmetry
\begin{equation}
\mathcal A = n_W+n_B-2n_{\rm match},
\label{eq:asym}
\end{equation}
where $n_W$ and $n_B$ are the numbers of surviving
White and Black pawns, and $n_{\rm match}$ is the
number of pawn squares that coincide after reflection.
Thus, $\mathcal A=0$ for perfectly symmetric pawn
structures, while larger values indicate greater
structural asymmetry.

We now combine this with other observables into a morphological
scorecard. For each position, we define the following
coordinates:
\begin{align}
X_{\rm commitment}
&=
\Lambda_W+\Lambda_B,
\\
X_{\rm center}
&=
A_W^{de}+A_B^{de}
-
A_W^w-A_B^w,
\\
X_{\rm activity}
&=
M_W+M_B,
\\
X_{\rm centralization}
&=
\kappa_W+\kappa_B,
\\
X_{\rm locked}
&=
B,
\\
X_{\rm tension}
&=
L,
\\
X_{\rm asymmetry}
&=
\mathcal A.
\end{align}

We additionally use the number of catalog names associated with a
position as a simple proxy for the density of opening theory around that
state. We call it \emph{theory burden}, with the important qualification
that name multiplicity is not literally the number of variations a
player must memorize.

The eight static coordinates have direct structural interpretations. Commitment measures the extent of irreversible pawn development, while center focus describes where this advancement is concentrated. Piece activity measures the overall involvement of the pieces, and centralization quantifies how strongly this activity is directed toward the center. Locked and tension characterize the degree of blockage and immediate interaction between opposing pawns, respectively. Asymmetry measures the difference between the White and Black pawn structures, while theory burden quantifies how densely a position is represented by named lines in the opening catalog.

\begin{figure*}[t]
\centering
\includegraphics[width=0.99\textwidth]{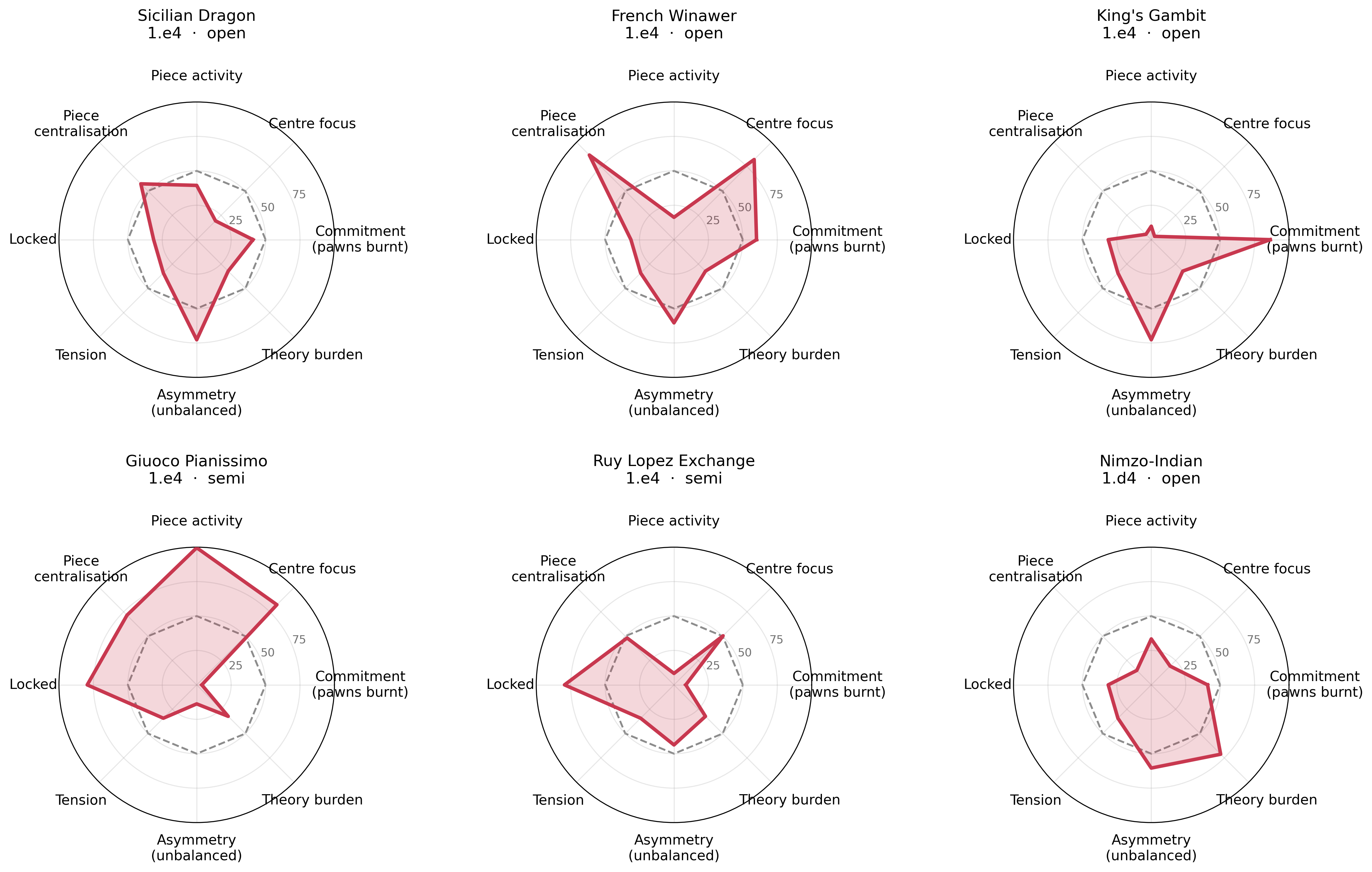}
\caption{
\textbf{Opening scorecards.}
Morphological profiles of six individual opening lines at $N=10$.
For each named opening family, the displayed line is the medoid of all
catalog lines sharing that prefix; no averaging is performed. Each
radial coordinate gives the percentile of the corresponding position
within the catalog, and the dashed circle denotes the catalog median.
The axes measure pawn commitment, central focus, piece activity and
centralization, pawn locking and tension, structural asymmetry, and theory
burden.
}
\label{fig:showcase}
\end{figure*}

Figure~\ref{fig:showcase} compares six individual opening lines at
$N=10$. For each named opening family, we select the medoid (the
actual catalog line with the smallest total distance to the other lines
of that family in morphological percentile space). The scorecards are
therefore not averages; each one corresponds to a single position reached
after ten plies.

The profiles reveal markedly different combinations of slow and fast
variables. The King's Gambit representative
(\texttt{1.e4 e5 2.f4 exf4 3.Bc4 Qh4+ 4.Kf1 d5 5.Bxd5 g5})
is the most strongly committed of the six examples, while its early piece
activity and centralization are both very low. The Giuoco Pianissimo
(\texttt{1.e4 e5 2.Nf3 Nc6 3.Bc4 Nf6 4.O-O Bc5 5.Nc3 d6})
shows almost the opposite profile: pawn commitment is minimal, yet piece
activity lies close to the maximum of the catalogue and is strongly
directed toward the center. These two positions illustrate directly that
pawn commitment does not determine the activity of the pieces organized
on top of it.

The French Winawer representative
(\texttt{1.e4 e6 2.d4 d5 3.Nc3 Bb4 4.Ne2 dxe4 5.a3 Be7})
provides a different combination. Its pawn commitment is already above
average, but its most distinctive feature is its strong central
organization: both center focus and piece centralization lie high in the
catalogue despite relatively low total piece activity. The Sicilian
Dragon representative
(\texttt{1.e4 c5 2.Nf3 Nc6 3.d4 cxd4 4.Nxd4 g6 5.Be3 Bg7}),
by contrast, is less centrally focused and considerably more asymmetric.
Thus the amount of piece activity and the spatial direction of that
activity again provide distinct information.

The Ruy Lopez Exchange representative
(\texttt{1.e4 e5 2.Nf3 Nc6 3.Bb5 a6 4.Bxc6 dxc6 5.O-O Qd6})
is comparatively restrained in both pawn commitment and piece activity,
while retaining a strongly locked pawn structure. The Nimzo-Indian
representative
(\texttt{1.d4 Nf6 2.c4 e6 3.Nc3 Bb4 4.Qc2 c5 5.dxc5 O-O})
combines intermediate pawn commitment with low piece centralization and
substantial asymmetry; among these examples it also stands out through its
comparatively large theory burden.

Taken together, these six concrete positions show that familiar opening
families can generate very different combinations of slow structural
organization and fast piece activity. No single coordinate orders them in
a natural way. The scorecards are therefore best viewed as compact
representations of the multidimensional morphology of individual opening
positions.

\section{Discussion and conclusions}
\label{sec:discussion}

\subsection{Main results and physical picture}

The main result of this work is that chess openings admit a natural
slow--fast decomposition. Pawn moves are irreversible and therefore
build a slowly evolving structural background that retains memory of earlier
decisions. Pieces, by contrast, can be redeployed much more rapidly on top of
this background. This separation provides a simple physical picture of an
opening as a combination of a persistent structural sector and faster
degrees of freedom.

The pawn sector also provides an effective coarse graining of opening
space. At $N=10$, the catalogue collapses from many named variations to
$619$ distinct positions and only $435$ joint pawn skeletons. The realized
pawn structures do not form an unstructured collection: they occupy a
largely connected and strongly modular network. This modular organization is
robust to retaining link direction and persists as the opening depth is
varied, although individual communities split and merge. Opening families
can therefore be viewed as regions of an evolving structural space rather
than as sharply separated classes.

The faster piece sector adds largely independent information. Pawn
commitment, spatial pawn organization, piece activity, and centralization
are only partially correlated, so openings cannot be arranged along a
single axis such as open--closed or quiet--sharp. Instead, they occupy a
genuinely multidimensional morphological space. This state-space
description complements the conventional opening tree: the tree records
how a position is reached, whereas the morphology characterizes the
structural and dynamical properties of the position that has been reached.

\subsection{Limitations and extensions}

It is worth mentioning that our approach has still several limitations. First, the attack field $m_c(s)$ is a geometric measure of piece influence rather than a complete
measure of tactical activity. In particular, it counts squares attacked by
a piece even when that piece is pinned, and it does not distinguish between
attacks with different tactical significance. A more detailed description
could incorporate exchanges, or position-dependent evaluations of square control.

Note that our analysis characterizes the morphological organization of
chess openings independently of how frequently they are played.
An interesting extension would be to use databases of actual
games to measure how players populate this morphological space,
identifying the most frequently visited regions and communities
of the pawn-skeleton network.


Second, the present analysis uses $N=10$ as its main snapshot and, in the
scorecards, focuses on White's slow and fast variables. Including the
corresponding Black variables would provide a more complete description of
individual positions. The analysis at $N=8,10,$ and $12$ already shows that
the coarse modular organization persists while individual communities
split and merge. A denser longitudinal analysis could follow these
transformations continuously and connect early pawn structures to more
mature middlegame formations.

Finally, the theory-burden coordinate is only a proxy for memorization.
The number of named catalogue lines reaching a position measures how
densely that position is represented in opening theory, but does not
directly quantify how much a player must know. The branching structure and
depth of plausible continuations would provide a more direct measure of
theoretical complexity.

\subsection{Navigating opening space}

Beyond the structural description itself, the morphological representation
suggests a practical use. Opening choice is traditionally organized through
historical families: a player first chooses, for example, the Sicilian,
French, or Ruy Lopez and then studies the variations that follow. The
present description proposes an inversion of this logic: a player could
instead specify the kinds of positions they prefer---for example, low pawn
commitment with active centralized pieces, strongly asymmetric pawn
structures, or positions with limited theory burden---and search the
catalogue for openings lying close to that target profile.

The morphological description can also reveal similarities between openings
that belong to different traditional families. Two openings with comparable
pawn commitment, centralization, asymmetry, or theory burden may lead to
similar types of positions despite having very different names and move
histories. This provides a complementary way to compare openings and may
help players identify alternatives that suit the same positional
preferences, or conversely build a repertoire that exposes them to
contrasting types of positions. Such an approach would complement rather
than replace traditional opening theory: opening names encode historical
lineages and move sequences, whereas morphology describes the kinds of
positions those sequences tend to produce.

More broadly, the traditional opening tree describes chess through
sequences of moves, while the present approach provides a complementary
description in terms of states and their geometry. Pawns form a slow,
history-bearing structural skeleton, whereas the pieces constitute faster
degrees of freedom that reorganize on top of it. Their combination defines
a multidimensional morphology of opening positions. 
Chess thus provides a particularly clear example of
slow--fast organization in a sequential complex system.
By describing openings through their pawn structures,
irreversible commitment, and piece activity, our approach
also provides a quantitative way to compare openings
and identify those leading to similar types of positions.

\begin{acknowledgments}

The analysis uses the open
\texttt{lichess-org/chess-openings} catalogue and the
\texttt{python-chess} software library.

\end{acknowledgments}

\bibliographystyle{apsrev4-2}
\bibliography{chess_biblio}

\end{document}